\documentstyle[12pt]{article}

\begin{document}

\title{ Recalculation of QCD Corrections
to  $b \to s \gamma$ Decay }

\author{ Chong-Shong Gao$^{a,b,c}$, Jing-Liang Hu$^{a,b}$,
	Cai-Dian L\"{u}$^{a,b,c}$\thanks{E-mail: lucd@itp.ac.cn},
 Zhao-Ming Qiu$^d$ \\
$a$ CCAST(World Laboratory), P.O.Box 8730, Beijing 100080, China\\
$b$ Physics Department, Peking University,
Beijing 100871, China\\
$c$ Institute of Theoretical Physics, Academia Sinica,\\
P.O.Box 2735, Beijing 100080, China\\
$d$ General Office, Chinese Academy of Sciences}

 \date{ 11 October 1993}
\maketitle

\begin{abstract}
We give a more complete calculation of $b \to s\gamma $ decay,
including leading log QCD corrections from $m_{top}$
to $M_W$ in addition to corrections  from $M_{W}$ to $m_b$.
We have included the full set of dimension-6 operators and corrected
numerical mistakes of anomalous dimensions in a previous paper\cite{Cho}.
Comparing with the calculations without QCD running from $m_{top}$
to $M_W$\cite{Mis}, the inclusive decay rate is found to be enhanced.
At $m_t=150$GeV, it results in 12\% enhancement, and for
$m_t=250$GeV, 15\% is found. The total QCD effect makes
an enhanced  factor of 4.2
at $m_t=150$GeV, and 3.2 for $m_t=250$GeV.
\end{abstract}
\bigskip


\newpage

		\section{Introduction}

Recently the CLEO collaboration has observed\cite{cleo} the
exclusive decay $B \rightarrow K^* \gamma $ with a branching
 fraction
of $(4.5 \pm 1.5 \pm 0.9) \times 10^{-5}$. A new upper limit
on the inclusive $b \rightarrow s \gamma$ process is also
obtained as
$B(b \rightarrow s \gamma)<5.4 \times 10^{-4}$ at 95\%
C.L.\cite{cleo2}.
This has been a subject of many papers\cite{Hew,Bar,Riz,Hew2,Ali,Bur}
recently. It has been argued that this experiment provides more
information about restrictions on
the Standard Model, Supersymmetry, Technicolor etc. Their
results are found to be sensitive to the theoretical
calculation of $b \to s \gamma$ decay.
In order to reduce the theoretical uncertainty,
a more accurate calculation of this decay rate is needed.

The radiative $b$ quark decay has already been calculated
 in several  papers\cite{Springer}--\cite{Mis}.
It is found to be strongly QCD-enhanced
(e.g. a factor of 7 for $m_t=80$ GeV and $\Lambda_{QCD}=300$ MeV
in ref.\cite{Mis} ). In other words, the strong interaction plays an
important
role in this decay. However, there are still some uncertainties in
these  papers.
In ref.\cite{Springer}, the
anomalous dimension matrix was truncated and the estimated uncertainty
due to this truncation is less than 15\%. Refs.\cite{Gri,Cel}
did not contain a full leading logarithmic analysis either, although they
included some of the terms neglected in ref.\cite{Springer}.\footnote{
See discussions in ref.\cite{Mis}.}
Ref.\cite{Mis} is believed to be a more accurate result.
But it did not include the QCD running from $m_{top}$
to $M_W$. Since the top quark is found to be heavier than W boson(
$m_{top} = 174\pm 10^{+13}_{-12}$ GeV.\cite{D0}),
a detailed calculation of this effect is needed .
Ref.\cite{Cho} does include this running, however there
are some errors in the calculation of anomalous dimensions, which can lead
to some changes in the final result.

In the present paper,
we recalculate the $b \to s \gamma$ decay including QCD running from
$m_{top}$ to $M_W$, and correct the errors in ref.\cite{Cho}, i.e.
its anomalous dimension matrix.
Furthermore, we use untruncated anomalous dimensions of QCD running from
$M_W$ to $m_b$.

\section{Matching at $\mu=m_t$}

In Minimal Standard Model, we first integrate
out the top quark, generating an effective five-quark theory.
By using the
renormalization group equation, we run the effective field theory
down to the W-scale, at which the weak bosons are removed. Finally
we continue running the effective field theory
down to b-quark scale to calculate the rate of radiative $b$ decay.
To maintain gauge invariance, we work in a background field
 gauge\cite{Abbott}.

The charged sector of Standard Model Lagrangian is
\begin{eqnarray}
 {\cal L}_{CC}&= &
  \frac{1}{\sqrt{2}} \mu^{\epsilon/2} g_2\left(\begin{array}{ccc}
    \overline{u} & \overline{c} & \overline{t} \end{array}\right)_L
\gamma_{\mu} V\left(\begin{array}{c} d \\ s \\ b \end{array}\right)_L
	  W_+^{\mu} \nonumber\\
        &  +&\frac{1}{\sqrt{2}} \frac{\mu^{\epsilon/2}g_2}{M_W}
      \left[\left(\begin{array}{ccc} \overline{u} & \overline{c} &
	      \overline{t} \end{array}\right)_R M_U V
	      \left(\begin{array}{c} d \\ s \\ b \end{array}\right)_L
	     -\left(\begin{array}{ccc} \overline{u} & \overline{c} &
	     \overline{t} \end{array}\right)_L V M_D
	    \left(\begin{array}{c} d \\ s \\ b \end{array}\right)_R
	     \right]\phi_+ \nonumber\\
	&  +&h.c..
\end{eqnarray}
Where V represents the $3 \times 3$ unitary Kobayashi-Maskawa matrix,
$M_U$ and $M_D$ denote the diagonalized quark mass matrices, the
subscript
$L$ and $R$ denote left-handed and right-handed quarks, respectively.

We first integrate out the top quark, introducing dimension-6
operators to include effects of the absent top quark.
In ref.\cite{Cho}, an approximation was made to keep only leading order
terms of $\delta = M_W^2/ m_t^2$, neglecting the next to leading
order charged current to W boson. To include the full set of
dimension-6 operators, we have to pick up five more operators involving
W bosons. Our operators now make a complete basis of dimension-6 operators.
Higher dimension operators are suppressed by factor $p^2/m_t^2$, where
$p^2$ characterizing the external momentum of b quark etc. $p^2\sim m_b^2$.
The basis operators are:
\begin{eqnarray}
O_{LR}^1  & =  &  -\frac{1}{16\pi^2} m_b \overline{s}_L D^2 b_R,
\nonumber\\
O_{LR}^2  &  =  &  \mu^{\epsilon/2} \frac{g_3}{16\pi^2}
	m_b \overline{s}_L \sigma^{\mu\nu} X^a b_R G_{\mu\nu}^a,
\nonumber\\
O_{LR}^3  &  =  &  \mu^{\epsilon/2} \frac{e Q_b}{16\pi^2}
	m_b \overline{s}_L \sigma^{\mu\nu} b_R	F_{\mu\nu},
\nonumber\\
Q_{LR}  &  =  &  \mu^{\epsilon} g_3^2 m_b
	\phi_{+}\phi_{-} \overline{s}_L b_R,
\nonumber\\
P_L^{1,A}  &  =  &  -\frac{i}{16\pi^2}  \overline{s}_L
  T_{\mu\nu\sigma}^A D^{\mu} D^{\nu} D^{\sigma} b_L,\nonumber\\
P_L^2 & = & \mu^{\epsilon/2} \frac{e Q_b}{16\pi^2}  \overline{s}_L
	\gamma^{\mu} b_L \partial^{\nu} F_{\mu\nu},\nonumber\\
P_L^3  &  =  &  \mu^{\epsilon/2} \frac{e Q_b}{16\pi^2}  F_{\mu\nu}
	\overline{s}_L 	\gamma^{\mu} D^{\nu}b_L,\nonumber\\
P_L^4  &  =  &  i \mu^{\epsilon/2} \frac{e Q_b}{16\pi^2}
	\tilde{F}_{\mu\nu}
	\overline{s}_L \gamma^{\mu} \gamma^5 D^{\nu} b_L,\nonumber\\
R_L^1 &  =  &  i \mu^{\epsilon} g_3^2 \phi_{+}\phi_{-} \overline{s}_L
	\not \!\! D b_L,\nonumber\\
R_L^2  &  =  &  i \mu^{\epsilon} g_3^2(D^{\sigma} \phi_+) \phi_{-}
	\overline{s}_L\gamma_{\sigma} b_L,
\nonumber\\
R_L^3  &  =  &  i \mu^{\epsilon} g_3^2 \phi_{+}(D^{\sigma} \phi_{-})
	\overline{s}_L\gamma_{\sigma} b_L,\nonumber\\
W_{LR} &  =  & -i \mu^{\epsilon} g_3^2 m_b W^{\nu}_{+}W_{-}^{\mu}
	\overline{s}_L \sigma_{\mu \nu} b_R, \nonumber\\
W_L^1  &  =  &  i \mu^{\epsilon} g_3^2 W^{\nu}_{+}W_{-}^{\mu}
	\overline{s}_L
	\gamma _{\mu}\not \!\! D \gamma _{\nu} b_L,\nonumber\\
W_L^2  &  =  &  i \mu^{\epsilon} g_3^2(D^{\sigma} W^{\nu}_+) W^{\mu}_{-}
	\overline{s}_L \gamma_{\mu} \gamma_{\sigma} \gamma_{\nu} b_L,
\nonumber\\
W_L^3  &  =  &  i \mu^{\epsilon} g_3^2 W_{+\mu} W^{\mu}_{-}
\overline{s}_L \stackrel{\leftrightarrow}{\not \!\! D} b_L, \nonumber\\
W_L^4  &  =  &  i \mu^{\epsilon} g_3^2 W^{\nu}_+ W^{\mu}_{-}
\overline{s}_L (\stackrel{\leftrightarrow}{D}_{\mu}\! \gamma_{\nu} +
	 \gamma_{\mu}\! \stackrel{\leftrightarrow}{D}_{\nu} ) b_L.
\end{eqnarray}
Where $\overline{s}_L\! \stackrel{\leftrightarrow}{D}_{\mu}
\!\gamma_{\nu} b_L$
stands for $\overline{s}_L D_{\mu} \gamma_{\nu} b_L +(D_{\mu}
 \overline{s}_L)
 \gamma_{\nu} b_L$ and the covariant derivative is defined as
$$D_{\mu}=\partial_{\mu}-i\mu^{\epsilon/2}g_3 X^a G_{\mu}^{a} -
i \mu^{\epsilon/2}eQ A_{\mu},$$
with $g_3$ denoting the QCD coupling constant.
The tensor $T_{\mu\nu\sigma}^A$ appearing in $P_L^{1,A}$
 assumes the following
Lorenz structure, the index $A$ ranging from 1 to 4:
\begin{equation}
\begin{array}{ll}
        T_{\mu\nu\sigma}^1=g_{\mu\nu} \gamma_{\sigma},
&T_{\mu\nu\sigma}^2=g_{\mu\sigma} \gamma_{\nu},\nonumber\\
	T_{\mu\nu\sigma}^3=g_{\nu\sigma} \gamma_{\mu},
&T_{\mu\nu\sigma}^4=-i \epsilon_{\mu\nu\sigma\tau}
	\gamma^{\tau} \gamma_5.
\end{array}
\end{equation}
Then we can write down our effective Hamiltonian
\begin{equation}
{\cal H}_{eff}=2 \sqrt{2} G_F V_{tb}V_{ts}^*\displaystyle \sum _i
C_i(\mu)O_i(\mu). \label{eff}
\end{equation}

The matching diagrams are displayed in Fig.1 and Fig.2. The diagrams
involving W bosons are introduced in addition to Goldstone
boson ones, So the number of diagrams considered now is 2--times bigger
than that in ref.\cite{Cho}. After tedious calculation we have
\begin{eqnarray}
C_{O_{LR}^1}&=&-\left(\frac{\frac{1}{2}+\frac{1}{2}\delta}
	{(1-\delta)^2}+\frac{\delta}{(1-\delta)^3}\log\delta\right),
\nonumber\\
C_{O_{LR}^2}&=&-\left(\frac{\frac{1}{2}}{(1-\delta)}+
	\frac{\frac{1}{2}\delta}{(1-\delta)^2}\log\delta\right),
\nonumber\\
C_{O_{LR}^3}&=& \left(\frac{1}{(1-\delta)}+
	\frac{\delta}{(1-\delta)^2}\log\delta\right),\nonumber\\
C_{P_L^{1,1}}&=& C_{P_L^{1,3}}\;\;=\;\;
	\left(\frac{\frac{11}{18}+\frac{5}{6}\delta
	-\frac{2}{3}\delta^2 +\frac{2}{9} \delta ^3}{(1-\delta)^3}+
\frac{\delta+\delta^2-\frac{5}{3}\delta^3 +\frac{2}{3} \delta ^4}
	{(1-\delta)^4}\log\delta\right),\nonumber\\
C_{P_L^{1,2}}&=&\left(\frac{-\frac{8}{9}-\frac{1}{6}\delta
	+\frac{17}{6}\delta^2 -\frac{7}{9} \delta ^3}{(1-\delta)^3}+
	\frac{-\delta+\frac{10}{3}\delta^3 -\frac{4}{3} \delta^4}
	{(1-\delta)^4}\log \delta \right),
\nonumber\\
C_{P_L^{1,4}}&=&\left(\frac{\frac{1}{2}-\delta
	-\frac{1}{2}\delta^2 +\delta^3}{(1-\delta)^3}+
\frac{\delta-3\delta^2+2\delta^3}{(1-\delta)^4}\log\delta\right),
\nonumber\\
C_{P_L^2}&=&\frac{1}{Q_b}\left(\frac{\frac{3}{4}+\frac{1}{2}\delta
	-\frac{7}{4}\delta^2 +\frac{1}{2} \delta^3 }{(1-\delta)^3}
	-\frac{1}{3} \delta
+\left(\frac{\frac{1}{6} +\frac{5}{6}\delta -\frac{5}{3}\delta^3+
	\frac{2}{3} \delta^4} {(1-\delta)^4}
	- \frac{1}{6} -\frac{1}{3} \delta \right)
	\log\delta \right),\nonumber\\
C_{P_L^3}&=&0,\nonumber\\
C_{P_L^4}&=&\frac{1}{Q_b}\left(\frac{-\frac{1}{2}-5\delta
	+\frac{17}{2}\delta^2 -3\delta^3 }{(1-\delta)^3}+
	\frac{-5\delta +7\delta^2 -2\delta^3}{(1-\delta)^4}\log\delta
	+4\delta \log\delta \right),
\nonumber\\
C_{R_L^1} &=& C_{R_L^2}\;\;=\;\;-C_{Q_{LR}}\;\;=
	\;\;1/g_3^2,\nonumber\\
C_{R_L^3} &=& 0,\nonumber\\
C_{W_{LR}} &=& C_{W_L^3}\;\;=\;\;C_{W_{L}^4}\;\;=0,\nonumber\\
C_{W_L^1} &=& C_{W_L^2}\;\;=\;\;\delta /g_3^2.
\end{eqnarray}

Notice that we have included the terms of all orders of $\delta =
M_W^2/m_t^2$ as far as the dimension-6 operators are concerned.
However, to really achieve higher accuracy in $\delta$, the dimension-8
operators should be considered in the matching of $M_W$ scale\footnote{We
thank Professor R. Barbieri for making this point explicit to us.}.
These coefficients are all from the finite part
integrations of electroweak loops. Terms like $\log(\mu^2/m_t^2)$
always accompanied by the infinity $1/\epsilon$
 vanishes here, because of our matching scale $\mu=m_t$. They will
be regenerated by renormalization group running of electroweak in the
next section.

\section{Running from $m_t$ to $M_W$}

The renormalization group equation satisfied by
the coefficient functions $C_i(\mu)$ is
\begin{equation}
\mu \frac{d}{d\mu} C_i(\mu)=\displaystyle\sum_{j}(\gamma^{\tau})_{
ij}C_j(\mu).\label{ren}
\end{equation}
Where the anomalous dimension matrix $\gamma_{ij}$
is calculated in practice by requiring
renormalization group equations for Green functions
with insertions of composite operators
to be satisfied order by order in perturbation theory.
Let $\Gamma_{O_i}^{(n)}$ denote a renormalized n--point
1PI Green function
with one insertion of operator $O_i$. Then the anomalous dimension
$\gamma_{ij}$ characterizing the mixing of operator $O_i$ into $O_j$
is determined from the renormalization group equation for
$\Gamma_{O_i}^{(n)}$,
\begin{equation}
\gamma_{ij} \Gamma_{O_j}^{(n)}=-\left(\mu \frac{\partial}{\partial
 \mu} +\beta \frac{\partial}{\partial g }+\gamma_m m \frac{\partial}
{\partial m} -n \gamma_{ext}\right) \Gamma_{O_i}^{(n)}.\label{renG}
\end{equation}
Here $\beta=\mu(d/d \mu )g$, $\gamma_m=(\mu/m)(d/d \mu)m$ and
 $n\gamma_{ext}$
stands for the wave-function anomalous dimensions arising from
 radiative
corrections to the Green function's $n$ external lines.

After evaluating the loop diagrams, we find the following
leading order weak mixing of operators, with the Q, R part
 agrees with ref.\cite{Cho}.
\begin{equation}
\begin{array}{rccl}
  \gamma=
   & \begin{array}{c} \\ Q_{LR}\\ R_L^1\\ R_L^2\\ R_L^3\\ W_{LR}\\
	 W_L^1\\	W_L^2\\ W_L^3\\ W_L^4  \end{array}
   & \begin{array}{c}
  	\begin{array}{cccccccc}
O_{LR}^1& O_{LR}^2& O_{LR}^3 & P_L^{1,A} & P_L^2 & P_L^3 & P_L^4 &
  	\end{array} \\

	\left(\begin{array}{ccccccccccccc}
	 0 && 0 &&& 0 && 0 &  0   & 0 && 0 \\
	 0 && 0 &&& 0 && 0 &  0   & 0 && 0 \\
         0 && 0 &&& 0 && 0 & -1/2 & 0 && 0 \\
         0 && 0 &&& 0 && 0 & 1/2  & 0 && 0 \\
	 0 && 0 &&& 6 && 0 &  0   & 0 && 0\\
	 0 && 0 &&& 0 && 0 &  0   & 0 && 12\\
	 0 && 0 &&& 0 && 0 & -1   & 0 && 0 \\
	 0 && 0 &&& 0 && 0 &  0   & 0 && 0 \\
	 0 && 0 &&& 0 && 0 &  0   & 0 && 0 \\
	\end{array}\right)

     \end{array}

   & 16\pi^2\; \displaystyle{ \frac{g_3^2}{8\pi^2} }.

\end{array}\label{weak}
\end{equation}
These mixing are all between operators induced by tree-diagram and
loop-diagram. The vanishing $\log(\mu^2/m_t^2)$ terms in the last section
are regenerated here by renormalization group equation.

The QCD anomalous dimensions for each of the operators
in our basis are
\begin{equation}
 \begin{array}{c}
     \begin{array}{ccccccccccccc}
	& && O_{LR}^1 & O_{LR}^2& O_{LR}^3& P_L^{1,1}& P_L^{1,2}&
	  P_L^{1,3}& P_L^{1,4}& P_L^{2}& P_L^{3}& P_L^{4}
     \end{array}\\
     \begin{array}{r}
  O_{LR}^1\\ O_{LR}^2\\ O_{LR}^3\\ P_L^{1,1}\\ \gamma=\; P_L^{1,2}\\
	  P_L^{1,3}\\ P_L^{1,4}\\ P_L^{2}\\ P_L^{3}\\ P_L^{4}
     \end{array}\left(\begin{array}{ccccccccccccccc}
	  \frac{20}{3} && 1 && -2 & 0 & 0 & 0 & 0 && 0 & 0 && 0 \\
 -8 && \frac{2}{3} && \frac{4}{3} & 0 & 0 & 0 & 0 && 0 & 0 && 0 \\
	 0 && 0 && \frac{16}{3} & 0 & 0 & 0 & 0 && 0 & 0 && 0 \\
	 6 && 2 && -1 & \frac{2}{3} & 2 & -2 & -2 && 0 & 0 && 0 \\
	 4 && \frac{3}{2} && 0 & -\frac{113}{36} & \frac{137}{18}
	 & -\frac{113}{36} &-\frac{4}{3} &&\frac{9}{4} & 0 && 0  \\
	 2 && 1 && 1 & -2 & 2 & \frac{2}{3} & -2 && 0 & 0 && 0 \\
	 0 && \frac{1}{2} && 2 & -\frac{113}{36} & \frac{89}{18}
	  & -\frac{113}{36} &
	 \frac{4}{3} && \frac{9}{4} & 0 && 0 \\
	 0 && 0 && 0 & 0 & 0 & 0 & 0 && 0 & 0 && 0 \\
	 0 && 0 && -\frac{4}{3} & 0 & 0 & 0 & 0 && 0 & 0 && 0 \\
	 0 && 0 && -\frac{4}{3} & 0 & 0 & 0 & 0 && 0 & 0 && 0
	\end{array}\right) \displaystyle{ \frac{g_3^2}{8\pi^2} },
\end{array}
\label{anom1}
\end{equation}

\begin{equation}
\begin{array}{cccc}
\gamma= & \begin{array}{c}
	\\ Q_{LR}\\ R_L^1\\ R_L^2\\ R_L^3\\ W_{LR}\\ W_L^1\\ W_L^2\\
		W_L^3\\ W_L^4\\
          \end{array}

	& \begin{array}{c}
	      \begin{array}{ccccccccc}Q_{LR} & R_L^1 & R_L^2 & R_L^3
			  & W_{LR} & W_L^1 & W_L^2 & W_L^3 & W_L^4
 		     \end{array}
	    \\
	      \left( \begin{array}{ccccccccccccc}
 \frac{23}{3} && 0  & 0  && 0 & 0 && 0 && 0 &  0  &      0\\
 0  && \frac{23}{3} & 0  && 0 & 0 && 0 && 0 &  0  &     0\\
 0  &&  0 & \frac{23}{3} && 0 & 0 && 0 && 0 &  0  &    0\\
 0  &&  0 & 0 && \frac{23}{3} & 0 && 0 && 0 &  0  &     0\\
 0  &&  0   &   0   &&  0    & 13 && 0 && 0  &  0  &     0\\
 0 && 0& 0 && 0 &-\frac{8}{3} &&\frac{23}{3} && 0 &-\frac{8}{9}
&\frac{16}{9}\\
 0  &&  0 & 0 && 0 & 0 && 0 && \frac{23}{3} &  0  &    0\\
 0  &&  0 & 0 && 0 & 0 && 0 && 0  & \frac{23}{3} &    0\\
 0  &&  0 & 0 && 0 & 0 && 0 && 0 & -\frac{16}{9} & \frac{101}{9}\\
	      \end{array} \right)
	  \end{array}
	& \displaystyle{ \frac{g_3^2}{8 \pi^2} }.

\end{array}\label{anom2}
\end{equation}

Comparing with ref.\cite{Cho}, except the $W$ part,
there are still some differences in the anomalous
dimension matrix, which may lie in omitting
a factor of 1/2 in ref.\cite{Cho} in calculating Feynman diagram like
Fig.3. After these changes, the whole matrix can be easily
diagonalized, and all eigenvalues are real, which is required to maintain
hermiticity of the effective Hamiltonian at all renormalization scales.
While in ref.\cite{Cho} it can not. In their case, some eigenvalues are
complex.

The solution to eqn.(\ref{ren}) appears in
obvious matrix notation as
\begin{equation}
C(\mu_2)=\left[\exp\int_{g_3(\mu_1)}^{g_3(\mu_2)}dg\frac
{\gamma^T(g)}{\beta(g)}\right] C(\mu_1).\label{solu}
\end{equation}
After inserting anomalous dimension (\ref{weak}--\ref{anom2}),
we can have the coefficients of operators at
$\mu=M_W$. And some of these operators
change a lot from ref.\cite{Cho} due to our improvements.
For details, see next section.

\section{Matching at $\mu=M_W$}

In order to continue running the basis operator coefficients down to
lower scales, one must integrate out the weak gauge bosons and
would-be
Goldstone bosons at $\mu=M_W$ scale. The diagrams are displayed in
Fig.4. In these new matching conditions, one finds the following
 relations
between coefficient functions just above and below $\mu=M_W$:
\begin{eqnarray}
C_{O_{LR}^1}(M_W^-)&=&C_{O_{LR}^1}(M_W^+), \nonumber\\
C_{O_{LR}^2}(M_W^-)&=&C_{O_{LR}^2}(M_W^+), \nonumber\\
C_{O_{LR}^3}(M_W^-)&=&C_{O_{LR}^3}(M_W^+),\nonumber \\
C_{P_L^{1,1}}(M_W^-)&=&C_{P_L^{1,1}}(M_W^+) + 2/9,\nonumber \\
C_{P_L^{1,2}}(M_W^-)&=&C_{P_L^{1,2}}(M_W^+) - 7/9, \nonumber\\
C_{P_L^{1,3}}(M_W^-)&=&C_{P_L^{1,3}}(M_W^+) + 2/9,\nonumber \\
C_{P_L^{1,4}}(M_W^-)&=&C_{P_L^{1,4}}(M_W^+) + 1, \nonumber\\
C_{P_L^2}(M_W^-)&=&C_{P_L^2}(M_W^+)
	-C_{W_L^2}(M_W^+) - 3/2, \nonumber\\
C_{P_L^3}(M_W^-)&=&C_{P_L^3}(M_W^+), \nonumber\\
C_{P_L^4}(M_W^-)&=&C_{P_L^4}(M_W^+) + 9.
\end{eqnarray}
In addition to these, there are new four-quark operators:
\begin{eqnarray}
O_1&=&(\overline{c}_{L\beta} \gamma^{\mu} b_{L\alpha})
	    (\overline{s}_{L\alpha} \gamma_{\mu} c_{L\beta}),
\nonumber\\
O_2&=&(\overline{c}_{L\alpha} \gamma^{\mu} b_{L\alpha})
	    (\overline{s}_{L\beta} \gamma_{\mu} c_{L\beta}),
\nonumber\\
O_3&=&(\overline{s}_{L\alpha} \gamma^{\mu} b_{L\alpha})
	    [(\overline{u}_{L\beta} \gamma_{\mu} u_{L\beta})+...+
	    (\overline{b}_{L\beta} \gamma_{\mu} b_{L\beta})],
\nonumber\\
O_4&=&(\overline{s}_{L\alpha} \gamma^{\mu} b_{L\beta})
	    [(\overline{u}_{L\beta} \gamma_{\mu} u_{L\alpha})+...+
	    (\overline{b}_{L\beta} \gamma_{\mu} b_{L\alpha})],
\nonumber\\
O_5&=&(\overline{s}_{L\alpha} \gamma^{\mu} b_{L\alpha})
	    [(\overline{u}_{R\beta} \gamma_{\mu} u_{R\beta})+...+
	    (\overline{b}_{R\beta} \gamma_{\mu} b_{R\beta})],
\nonumber\\
O_6&=&(\overline{s}_{L\alpha} \gamma^{\mu} b_{L\beta})
	    [(\overline{u}_{R\beta} \gamma_{\mu} u_{R\alpha})+...+
	    (\overline{b}_{R\beta} \gamma_{\mu} b_{R\alpha})],
\end{eqnarray}
with coefficients
$$ C_i(M_W)=0, \;\; i=1,3,4,5,6, \;\; C_2(M_W)=1.$$

     To simplify the calculation and compare with the previous results,
equations of motion(EOM)\cite{eom} is used to reduce all the remaining
two-quark operators to the gluon and photon magnetic moment
operators $O_{LR}^2$ and $O_{LR}^3$.
The  effective Hamiltonian then appears just below the W-scale as
\begin{eqnarray}
{\cal H}_{eff} && =\frac{4G_F}{\sqrt{2}} V_{tb}V_{ts}^*
	\displaystyle \sum_{i}
                C_i(M_W^-) O_i(M_W^-)\nonumber\\
	& &\stackrel{EOM}{\rightarrow}
		\frac{4G_F}{\sqrt{2}} V_{tb}V_{ts}^*\left\{
		\left(-\frac{1}{2}C_{O_{LR}^1}
	+C_{O_{LR}^2}-\frac{1}{2}C_{P_L^{1,1}}-\frac{1}{4}C_{P_L^{1,2}}
	+\frac{1}{4}C_{P_L^{1,4}}\right) O_{LR}^2\right.\nonumber\\
	&&\;\;\;\;\;\;+\left(-\frac{1}{2}C_{O_{LR}^1}
	+C_{O_{LR}^3}-\frac{1}{2}C_{P_L^{1,1}}-\frac{1}{4}C_{P_L^{1,2}}
		+\frac{1}{4}C_{P_L^{1,4}}-\frac{1}{4}C_{P_L^3}
		-\frac{1}{4}C_{P_L^4}\right)O_{LR}^3\nonumber\\
	&&\;\;\;\;\;\;\left.+({\rm four-quark\;\, operators})\right\}.
\end{eqnarray}

For completeness, we first give the explicit expressions of
the coefficient of operator $O_{LR}^2$ and
$O_{LR}^3$ at $\mu=M_W^-$,
\nopagebreak[1]
\begin{eqnarray}
C_{O_{LR}^2}(M_W^-) &= & \left( \frac{\alpha _s (m_t)} {\alpha _s (M_W)}
	\right) ^{ \frac{14}{23} } \left\{ -\frac{1}{2}C_{O_{LR}^1}(m_t)
+C_{O_{LR}^2}(m_t) -\frac{1}{2}C_{P_L^{1,1}}(m_t) \right.\nonumber\\
&&	\;\;\;\;\;\;\;\;\;\;\;\;\;\;\;\;\;\;
	\left. -\frac{1}{4}C_{P_L^{1,2}}(m_t)
	+\frac{1}{4}C_{P_L^{1,4}}(m_t)\right\}
	+\frac{1}{3} ,\label{c2}
\end{eqnarray}
\begin{eqnarray}
&\displaystyle{
C_{O_{LR}^3}(M_W^-) = \left( \frac{\alpha _s (m_t)} {\alpha _s (M_W)}
	\right) ^{ \frac{16}{23} } \left\{ C_{O_{LR}^3}(m_t)
	+8 C_{O_{LR}^2}(m_t) \left[1-\left( \frac{\alpha _s (M_W)}
{\alpha _s (m_t)} \right) ^{ \frac{2}{23} } \right]\right.}&\nonumber\\
&\displaystyle{	+\left[-\frac{9}{2} C_{O_{LR}^1}(m_t)
	-\frac{9}{2}C_{P_L^{1,1}}(m_t)
-\frac{9}{4}C_{P_L^{1,2}}(m_t) +\frac{9}{4}C_{P_L^{1,4}}(m_t)\right]
\left[1- \frac{8}{9} \left( \frac{\alpha _s (M_W)}
{\alpha _s (m_t)} \right) ^{ \frac{2}{23} } \right] }&\nonumber\\
&\displaystyle{	\;\;\;\left.-\frac{1}{4}C_{P_L^4}(m_t)
+\frac{9}{23} 16\pi^2 C_{W_L^1}(m_t) \left[1- \frac{\alpha _s (m_t)}
	{\alpha _s (M_W)} \right] \right \}
	-\frac{23}{12} }. &  \label{c3}
\end{eqnarray}
They are expressed by coefficients at $\mu=m_t$ and QCD coupling
$\alpha_s$.
So it is convenient to utilize these formula.

In the previous paper\cite{Cho}, the higher order
terms of $M_W^2/m_t^2$ are included in this stage by
hands in order to match the previous work\cite{Springer}--\cite{Mis}
when $m_t \to M_W$. Therefore it is unnatural.
While in this work, we keep higher order terms of
$M_W^2/m_t^2$ from the very beginning at $\mu =m_t$ scale. If
the QCD correction is ignored (by setting $\alpha_s(m_t)=\alpha_s(M_W)$
 in eqn.(\ref{c2}),(\ref{c3}) ), our results would reduce to
the previous results\cite{Springer,Mis} exactly
where the top quark and W bosons are integrated out together.
This is required by the correctness of the effective Hamiltonian, and
it is also a consistent check.

If we rewrite our operators $O_{LR}^3$, $O_{LR}^2$
as $O_7$, $O_8$ like ref.\cite{Mis},
\begin{eqnarray}
O_7&=&(e/16\pi^2) m_b \overline{s}_L \sigma^{\mu\nu}
	    b_{R} F_{\mu\nu},\nonumber\\
O_8&=&(g/16\pi^2) m_b \overline{s}_{L} \sigma^{\mu\nu}
	    T^a b_{R} G_{\mu\nu}^a.
\end{eqnarray}
then
\begin{eqnarray}
C_7(M_W) &=& \frac{1}{3} C_{O_{LR}^3}(M_W^-), \nonumber\\
C_8(M_W) &=& - C_{O_{LR}^2}(M_W^-).
\end{eqnarray}
The obvious differences from QCD correction to $C_7(M_W)$ and
$C_8(M_W)$
can easily be seen from Fig.5 and Fig.6.
In comparison to ref.\cite{Cho}, the enhancement of
coefficient of operator $O_7$ is almost  the same size,
but the values for $O_8$ are quite different. Here the effect to
$O_8$  is an enhancement rather than a suppression as in
ref.\cite{Cho}. At $m_t=150GeV$, it is enhanced a factor of 40\%
in comparison to ref.\cite{Cho}.
These changes come mainly from the corrections
of anomalous dimensions described earlier. Since $C_7(M_W)$
and $C_8(M_W)$ are both the input of the following QCD
running from $M_W$ to $m_b$,
It will be is expected to change the final result.

\section{The $\overline{B} \rightarrow X_s \gamma$ decay rate}

The running of
the coefficients of operators from $\mu=M_W$ to $\mu=m_b$
was well described in ref.\cite{Mis}. After this running we have
the coefficients of operators at $\mu=m_b$ scale. Table 1 gives
the numerical
values of coefficients of operators $O_7$, $O_8$ with different
input of
top quark
mass. Here we use $M_W=80.22$GeV, $m_b=4.9$GeV. Both $C_7(m_b)$ and
$C_8(m_b)$ are enhanced in comparison to values obtained by Misiak
\cite{Mis}.
The obtained values of
$C_6$ at the $m_b$ scale are -0.030, -0.036, -0.041 for
$\Lambda_{QCD}^{f=5}=100$, 200 and 300 MeV, respectively. The
corresponding values of $C_5$ are 0.007, 0.008 and 0.009. They
 are just
the same as ref.\cite{Mis}.

The leading order $b \rightarrow s\gamma$ matrix element of $H_{eff}$
is given by the sum of operators $O_5$, $O_6$ and $O_7$,
\begin{equation}
<H_{eff}>=-2 \sqrt{2} G_F V_{ts}^* V_{tb} \left\{ C_7(\mu)+Q_d
[C_5(\mu)+3 C_6
(\mu)] \right\} <|O_7|>. \label{Heff}
\end{equation}
Therefore, the sought amplitude will be proportional to the squared
 modulus of
\begin{equation}
	C_7^{eff}(m_b)=C_7(m_b)+Q_d\;[C_5(m_b)+3 C_6(m_b)] \label{C7}
\end{equation}
instead of $|C_7(M_b)|^2$ itself.

Following ref.\cite{Springer}--\cite{Mis},
\begin{equation}
BR(\overline{B} \rightarrow X_s \gamma) /BR(\overline{B}
\rightarrow X_c e\overline{\nu}) \simeq\Gamma(b\rightarrow
s\gamma)/\Gamma
(b\rightarrow ce\overline{\nu}).
\end{equation}
Then applying eqs.(\ref{Heff}),(\ref{C7}), one finds
\begin{equation}
\frac{BR(\overline{B} \rightarrow X_s \gamma)}{BR(\overline{B}
\rightarrow X_c e \overline{\nu})} \simeq \frac{6 \alpha_{QED}}{\pi
 g (m_c/m_b)}
|C_7^{eff}(m_b)|^2 \left(1-\frac{2 \alpha_{s}(m_b)}{3 \pi} f(m_c/m_b)
\right)^
{-1},
\end{equation}
where $g(m_c/m_b)\simeq 0.45$ and $f(m_c/m_b)\simeq 2.4$ corresponding
to the phase space
factor and the one-loop QCD correction to the semileptonic decay,
respectively\cite{Cabi}. The electromagnetic fine structure constant
evaluated at the $b$ quark scale takes value as $\alpha_{QED}(m_b)=
1/132.7$.
 Afterwards one obtains the $\overline{B} \rightarrow X_s
\gamma$ decay rate normalized to the quite well established
 semileptonic
decay rate. The results are summarized in Fig.7 as functions of
 the top quark
mass: The two upper solid lines represent the QCD-corrected ratio
 of the decay
rates, corresponding to $\Lambda_{QCD}^{f=5}=100$ MeV and
$\Lambda_{QCD}^{f=5}=300$ MeV, respectively. While the dashed lines
 correspond to the results obtained by Misiak\cite{Mis}. The
QCD-uncorrected values are also shown.

In this figure, we can easily see that, at $m_t=150$GeV, it results
in 12\% enhancement from Misiak's result\cite{Mis},
and for $m_t=250$GeV, 15\% is found.

\section{Conclusion}

As a conclusion, we have given the full leading log QCD
corrections(include
QCD running from $m_{top}$ to $M_W$), with whole anomalous dimension
matrix untruncated. Comparison to the previous calculation\cite{Cho},
three points are improved:

(1) We have included the full set of dimension-6 operators.

(2) We correct errors of anomalous dimensions in ref.\cite{Cho}.

(3) We use untruncated anomalous dimensions in QCD running from
$M_W$ to $m_b$ instead of truncated ones.

In fact, point(2) makes an enhancement while point(3) leads to a
suppression. For point(1) there is no significance change in final
result. The total result does not change a lot, e.g. a suppression
of 4\% comparing ref.\cite{Cho}. Table~2 gives the results
by different authors.

The whole QCD-enhancement of
the  $BR(\overline{B} \rightarrow X_s \gamma)$ makes a factor of 4.4
at $m_t=145$GeV, and 3.2 at $m_t=250$GeV, when $\Lambda_{QCD}=200$MeV.

The branching ratio $BR(\overline{B} \rightarrow X_s \gamma)/
BR(\overline{B} \rightarrow X_c e \overline{\nu})$ ranges from
about $3\times10^{-3}$ to
$6\times10^{-3}$ as $m_t$ varies from 90GeV to
250GeV($\Lambda_{QCD}=200$MeV).
Although this result is not quite different from
the previous calculations, our improvements lie in reducing some
theoretical uncertainties. This improvements are important, since
the anomalous dimension matrix of ref.\cite{Cho} is not a hermitian
one, and its results are not faithful.

The gluon magnetic moment operator $O_8$ is also enhanced from
Misiak's
result\cite{Mis}, which is not the present paper's interest.

\bigskip
\bigskip
{\noindent \bf {\Large Acknowledgement}}
\bigskip

One of the authors(C.D. L\"{u}) is grateful to Prof. Xiaoyuan Li for
stimulating of this work and helpful discussions.
We also thank Prof. Y.B. Dai, T. Huang, Y.P. Kuang, Z.X. Zhang,
G.D. Chao, and Dr. Y.Q. Chen, Y. Liao, Q. Wang, Q.H. Zhang for
helpful discussions. This work is partly supported by the National
Natural Science Foundation of China
 and the Doctoral Program Foundation of Institution of Higher
Education.


\begin{table}
\begin{center}
\caption{Numerical results for coefficients of operators
$C_i(m_b)$.}
\begin{tabular}{|c|ccc|ccc|}
\hline
$m_{top}$(GeV) & \multicolumn{3}{|c|}{$c_7(m_b)$} &
				\multicolumn{3}{|c|}{$c_8(m_b)$ }\\
\hline
   & $\Lambda$=100  &$\Lambda$=200  &$\Lambda$=300  &$\Lambda$=100
		&$\Lambda$=200    & $\Lambda$=300\\
\hline
100    &  -0.297  & -0.324   & -0.345  & -0.151  & -0.163  & -0.173\\
\hline
110    &  -0.307  & -0.333   & -0.353  & -0.155  & -0.168  & -0.177\\
\hline
120    &  -0.315  & -0.341   & -0.361  & -0.160  & -0.172  & -0.181\\
\hline
130    &  -0.323  & -0.349   & -0.368  & -0.163  & -0.175  & -0.184\\
\hline
140    &  -0.330  & -0.356   & -0.375  & -0.166  & -0.178  & -0.187\\
\hline
150    &  -0.337  & -0.362   & -0.381  & -0.169  & -0.181  & -0.190\\
\hline
160    &  -0.343  & -0.368   & -0.387  & -0.172  & -0.184  & -0.193\\
\hline
170    &  -0.349  & -0.374   & -0.393  & -0.174  & -0.186  & -0.195\\
\hline
180    &  -0.354  & -0.379   & -0.398  & -0.176  & -0.188  & -0.197\\
\hline
190    &  -0.360  & -0.384   & -0.402  & -0.178  & -0.190  & -0.199\\
\hline
200    &  -0.364  & -0.388   & -0.407  & -0.180  & -0.191  & -0.200\\
\hline
210    &  -0.369  & -0.393   & -0.411  & -0.181  & -0.193  & -0.202\\
\hline
220    &  -0.373  & -0.397   & -0.415  & -0.183  & -0.194  & -0.203\\
\hline
230    &  -0.377  & -0.401   & -0.419  & -0.184  & -0.196  & -0.204\\
\hline
240    &  -0.381  & -0.404   & -0.422  & -0.185  & -0.197  & -0.205\\
\hline
250    &  -0.384  & -0.408   & -0.426  & -0.187  & -0.198  & -0.207\\
\hline
\end{tabular}
\end{center}
\end{table}

\begin{table}
\begin{center}
\caption{Enhancement(+) or suppression(--) of $b \to s\gamma$ decay rates
obtained by Cho et al.[14], Misiak[12] and the present
improved ones relative to earlier result by Grinstein
et al.[9] }
\begin{tabular}{|c|c|c|c|c|}
\hline
	& Grinstein et al. & Cho et al. & Misiak & This paper\\
\hline
QCD running $ M_W \to m_b$ & truncated & truncated & untruncated &
untruncated\\
\hline
QCD running $m_t \to M_W$ & neglected & performed & neglected & performed\\
\hline
$m_{top}$=150GeV  & 1  & +8\%  & --6\% & +5\% \\
\hline
$m_{top}$=250GeV  &  1  & +14\%  & --5\% & +10\% \\
\hline
\end{tabular}\label{comp}
\end{center}
\end{table}

\parindent=0pt
{ \bf {\Large Figure Captions}}
\bigskip

Fig.1 Leading order matching conditions at the top quark scale for the
1PI Green functions  in the full Standard Model and in the intermediate
effective field theory.

Fig.2 One loop matching conditions at the top quark scale for the
1PI Green functions  in the full Standard Model and in the intermediate
effective field theory.

Fig.3 One of the Feynman diagram in calculating Anomalous dimensions,
with the heavy dot denoting high dimension operator.

Fig.4 Matching conditions at $\mu=M_W$ for four quarks and two
quarks 1PI Green functions in the intermediate
effective field theory and effective field theory below W scale.

Fig.5 The photon magnetic moment operator's coefficient $C_7(M_W)$ for
different top quark mass. The ones with and without QCD corrections
are indicated by solid and dashed lines respectively. ($\Lambda=300$MeV
is used)

Fig.6 The gluon magnetic moment operator's coefficient $C_8(M_W)$ for
different top quark mass. The ones with and without QCD corrections
are indicated by solid and dashed lines respectively. ($\Lambda=300$MeV
is used)

Fig.7 BR($\overline{B} \rightarrow x_s \gamma$) normalized to
	BR($\overline{B} \rightarrow x_c e \overline{\nu}$),
	as function of top quark mass. The upper solid lines indicated
	our results for a full QCD correction.
	Dashed lines correspond to Misiak's results without QCD running
	from $m_{top}$ to $ M_W$.

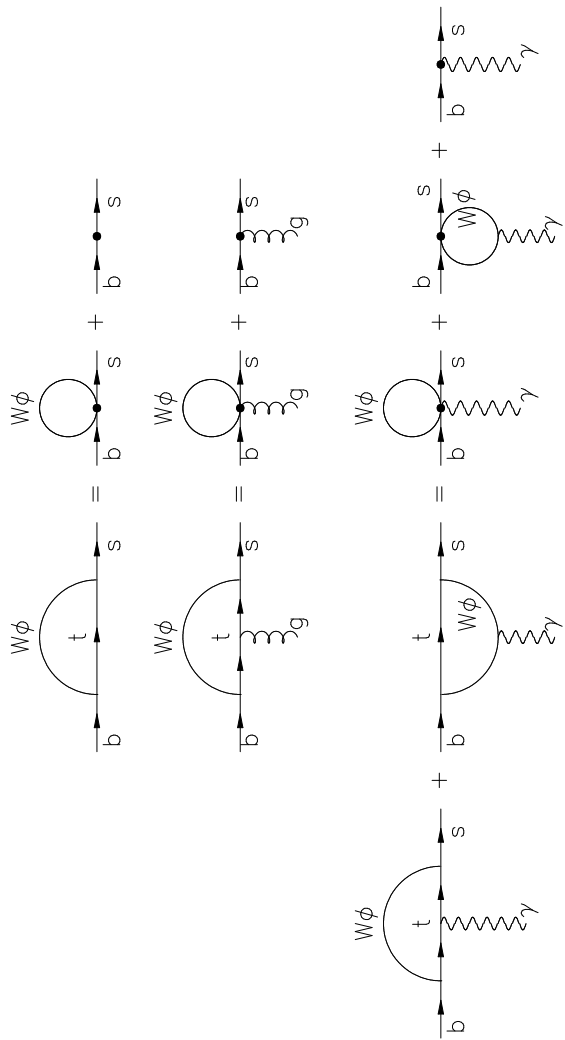
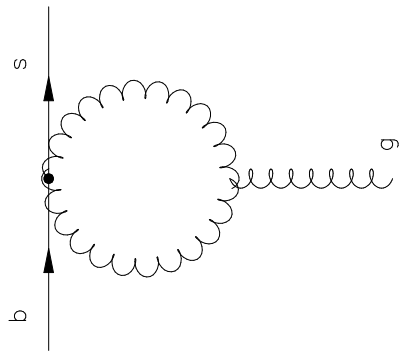
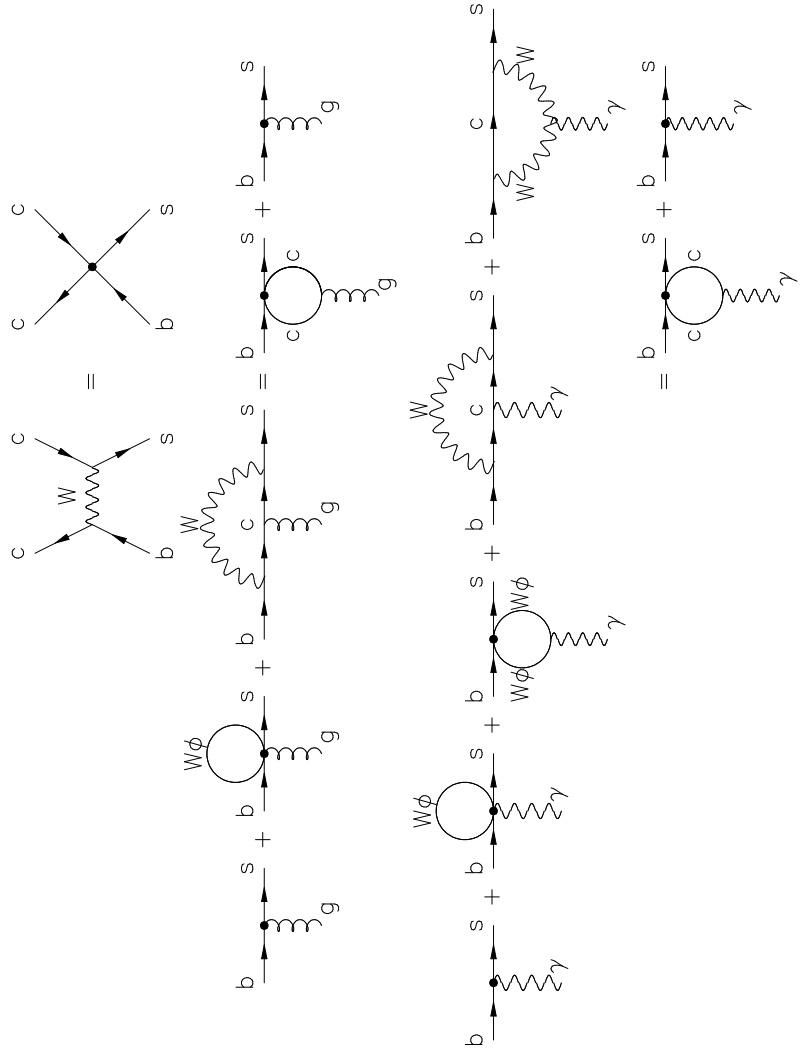
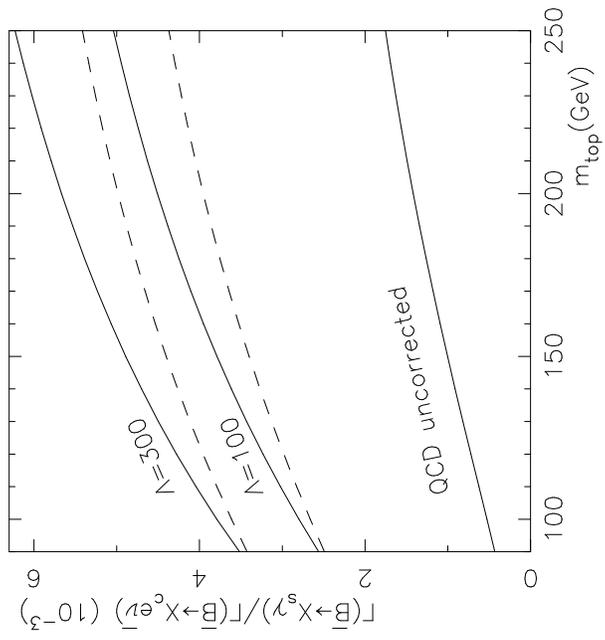
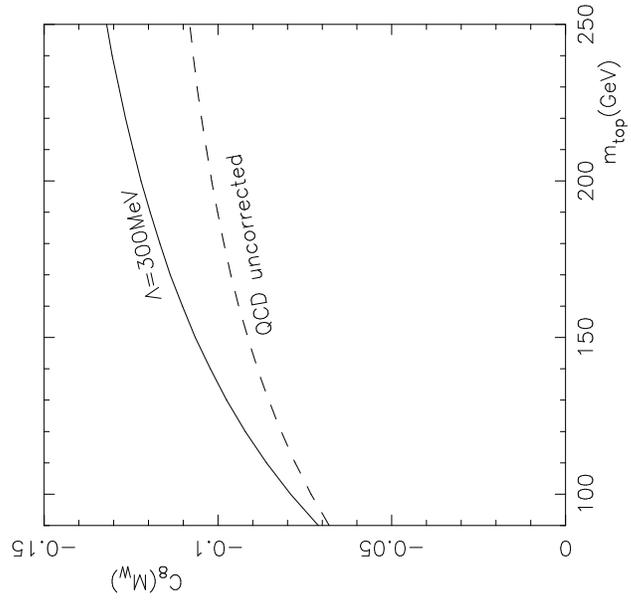
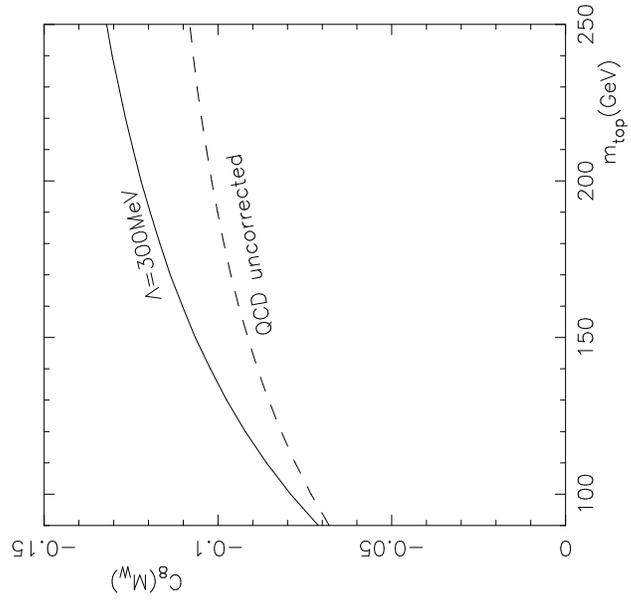
\end{document}